\documentclass[aps,prd,a4paper,showpacs,12pt]{article}

\usepackage{multicol}
\usepackage{subfigure}
\usepackage {graphicx}
\usepackage{color}
\usepackage{xcolor}
\usepackage{threeparttable}
\usepackage{amsfonts}
\usepackage{times}
\usepackage{setspace}
\usepackage{balance}
\usepackage{lastpage}
\usepackage{amsthm}
\usepackage[tbtags]{amsmath}
\usepackage{nomencl}
\usepackage{pstricks}
\usepackage{pstricks-add}
\usepackage{pst-plot,pstricks-add}

\usepackage{amsmath,amsfonts,amssymb,simplewick}
\usepackage{float}

\usepackage{amsmath,amssymb,amsfonts,epsfig,graphicx,euscript}%

\usepackage[bookmarks,bookmarksnumbered,linktocpage,pdfstartview=FitH]{hyperref} 
\hypersetup{colorlinks,%
	citecolor=red,%
	filecolor=blue,%
	linkcolor=blue,%
	urlcolor=blue,%
	pdftex}
\usepackage{amsmath}
\usepackage{amsfonts}
\usepackage{graphicx}
\usepackage{caption}
\usepackage{float}
\usepackage[all]{hypcap}
\numberwithin{equation}{section}
\usepackage{cite}
\usepackage{calc}
\newlength{\spacer}
\newsavebox{\mybox}

\newcommand{\bse}{\begin{subequations}}
	\newcommand{\ese}{\end{subequations}}
\newcommand{\be}{\begin{equation}}
\newcommand{\ee}{\end{equation}}
\newcommand{\bea}{\begin{eqnarray}}
\newcommand{\eea}{\end{eqnarray}}
\newcommand{\ba}{\begin{array}}
	\newcommand{\ea}{\end{array}}

\renewcommand{\thefootnote}{\fnsymbol{footnote}}

\begin{document}

	\begin{center}
		{ \large{\textbf{Contribution of the chiral vortical effect to the evolution of the hypermagnetic field and the matter-antimatter asymmetry in the early Universe}}} 
		\vspace*{1.5cm}
		\begin{center}
	{\bf S. Abbaslu\footnote{s$_{-}$abbasluo@sbu.ac.ir}$^1$, S. Rostam Zadeh\footnote{sh$_{-}$rostamzadeh@ipm.ir}$^2$ and S. S. Gousheh\footnote{ss-gousheh@sbu.ac.ir}$^1$}\\%
\vspace*{0.5cm}
{\it{$^1$Department of Physics, Shahid Beheshti University, G.C., Evin, Tehran 19839, Iran\\$^2$School of Particles and Accelerators, Institute for Research in Fundamental Sciences (IPM), P.O.Box 19395-5531, Tehran, Iran}}\\

			\vspace*{1cm}
		\end{center}
	\end{center}
	\begin{center}
		\today
	\end{center}
	
	\renewcommand*{\thefootnote}{\arabic{footnote}}
	\setcounter{footnote}{0}
	
	
	\begin{center}
		\textbf{Abstract}
	\end{center}

In this paper, we study the contribution of the chiral vortical effect, in addition to that of the chiral magnetic effect, to the evolution of the hypermagnetic field and the matter-antimatter asymmetry in the symmetric phase of the early Universe in the temperature range $100 \mbox{GeV}\leq T \leq 10\mbox{TeV}$. We choose a fully helical Chern-Simons wave configuration for the velocity and the hypermagnetic vector potential fields. The latter makes the plasma force-free in the absence of  viscosity. We show that the most pronounced effect of the chiral vorticity is the production and initial growth of the hypermagnetic field. In particular, we show that in the presence of a non-zero matter asymmetry, the hypermagnetic field can grow from zero initial value only in the presence of a non-zero vorticity field. Moreover, we show that larger initial growths not only result in larger maximum values of the hypermagnetic field, but also cause the saturation of the hypermagnetic field and the conversion of the lepton-baryon asymmetry to occur more quickly, i.e., at a higher temperature. We show that the damping of the vorticity due to the presence of viscosity, which typically occurs extremely rapidly, does not significantly affect the evolution.
\newpage

\section{INTRODUCTION}
The origin of the matter-antimatter asymmetry in the Universe is an open problem in the particle physics and cosmology. The amplitude of the baryon asymmetry of the Universe (BAU) has been obtained from the observations of the cosmic microwave background (CMB) and  the big bang nucleosynthesis (BBN) and the current accepted estimate is $\eta_{B}\sim 10^{-10}$\cite{{bas1},{bas2},{bas3}}. Many different mechanisms have been suggested for producing this asymmetry from an initial matter-antimatter symmetric state \cite{lepto,lepto2,leptobaryo}. Assuming the CPT invariance, Sakharov stated three necessary conditions for generating the BAU (baryogenesis): i) baryon number violation, ii) C and CP violation, iii) departure from thermal equilibrium \cite{Sakharov1}. However, it has been shown that the third condition is not necessary in the absence of the CPT invariance \cite{{add1},{bertolami}}. These conditions can be satisfied within the standard model of particle physics. Charge conjugation symmetry is violated in the weak interactions and CP is slightly violated through the CKM mechanism. The departure from thermal equilibrium can occur due to phase transitions and the expansion of the Universe. Furthermore, the baryon and lepton numbers are violated due to the weak sphaleron processes and the hypercharge Abelian anomaly at finite temperature \cite{gth,matter}. In fact, the matter-antimatter asymmetry generation and the magnetogenesis, which is the generation of long range magnetic fields in the Universe, are strongly intertwined via these hypercharge Abelian anomalous effects. 

Long range magnetic fields have been widely observed in galaxies, superclusters, and recently in the intergalactic medium (IGM). The strength of these magnetic fields have been measured or estimated by applying different methods \cite{{1},{11-1},{3}}. The induced Faraday rotation effect is used to measure the strength of the galactic magnetic fields, which is of the order of the microgauss in the Milky Way and several spiral galaxies \cite{11-1}. The temperature anisotropy of CMB puts an upper bound on the strength of the magnetic fields, $B\le10^{-9} \mbox{G}$ on the CMB scales $\lambda\simeq1$Mpc \cite{11}. The observations of the gamma rays from blazars put the strength of the intergalactic magnetic fields (IGMFs) in the range $B\simeq 10^{-17}-3\times10^{-14}$G on the scales as large as $\lambda\simeq1$Mpc \cite{Ando,Essey,Chen}. Furthermore, a non-vanishing helicity of these magnetic fields, with the strength $B\simeq5.5\times 10^{-14}$G, has been inferred\cite{Chen2}.


The origin of these galactic and intergalactic magnetic fields is also an open problem \cite{{4}, {5}, {6}}. Different mechanisms have been suggested which generally pursue one of the following two approaches to explain the origin and evolution of these long range magnetic fields in the Universe. One approach investigates the generation of the magnetic fields through different astrophysical mechanisms by assuming that the initial weak magnetic field is produced via a battery mechanism \cite{{7}, {8}, {9}}. The other one assumes that the magnetic fields have cosmological origin, that is, the present magnetic fields are produced from seed fields in the early Universe \cite{{10},{10-1},{10-2},{10-3},{10-4}}. Indeed, the presence of the magnetic fields at high redshifts everywhere in the Universe reinforces the idea that they have cosmological origin \cite{10-4}. The calculations of magnetic fields produced after the inflation usually suffer from the small-scale problem, that is, their comoving correlation length is much smaller than the observed scales of the magnetic fields in the Universe. In this paper, we pursue the latter approach and present a model that can produce hypermagnetic fields before the electroweak phase transition (EWPT). 


There are different processes that influence the evolution of the magnetic fields, such as the adiabatic expansion, the Abelian anomalous effects, the magnetohydrodynamics turbulent dynamo effect, the viscosity diffusion, the inverse cascade, and the direct cascade. Among these, the Abelian anomalous effects
are prominent since, as mentioned before, they interconnect the evolution of the magnetic fields to that of the matter-antimatter asymmetries \cite{{26},{27},{28},{30},{31},{32},{33},{34},{shiva3}}. These anomalous effects show up through the Abelian anomaly and the Abelian Chern-Simons term.
In addition to the Abelian anomalous effects, the magnetohydrodynamics turbulent dynamo effect also influences the evolution of the cosmological magnetic fields and their correlation length \cite{{18},{19},{20},{21},{22},{23},{24},{25}}. Indeed, turbulence is a complex phenomenon, and one of the characteristic parameters in the turbulence is the Reynolds number $Re=Lv/\nu$, where $L$ is the characteristic length scale, $v$ is the velocity, and $\nu$ is the kinematic viscosity. In this paper, we do not take turbulence into account; however, we study the Abelian anomalous effects, while taking into account the effects of velocity, as manifested in the form of vorticity, and the viscosity in the plasma.

The Ablelian gauge fields, unlike the non-Abelian ones which acquire mass gap $\sim g^{2}T$, remain massless. This, together with the fact that the plasma has high conductivity, make it possible only for the Abelian magnetic fields to survive in the plasma as long range gauge fields.
The Abelian $U_Y(1)$ anomaly emerges as a result of the chiral coupling of the Abelian hypercharge gauge fields to the fermions in the symmetric phase. The Abelian $U_Y(1)$ anomalous processes violate the baryon number $B$ and the lepton number $L$, but preserve $N_{i}=B/n_{G}-L_{i}$ and consequently $B-L$. Here, $L_{i}$ is the lepton number of the $i$th generation and $n_{G}$ is the number of generations. Indeed, these charges are the well known conserved charges of the Standard Model which, along with the ones discussed below, can be used to describe the plasma in thermal equilibrium.

In thermal equilibrium, the electroweak plasma can be described by $n_{G}$ chemical potentials $\mu_{i}$, corresponding to the aforementioned conserved charges, where $i=1,..,n_{G}$. Furthermore, due to the hypercharge neutrality of the electroweak plasma, there is also another chemical potential $\mu_{Y}$ which corresponds to the hypercharge of the plasma. Moreover, at temperatures higher than $T_{RL}\sim 10 \mbox{TeV}$, the right-handed electron chirality flip rate is much lower than the Hubble expansion rate. Therefore, the right-handed electron chirality flip processes are out of thermal equilibrium \cite{{16},{17}}, and in the absence of the Abelian anomaly, the number of right-handed electrons is perturbatively conserved as well. Therefore, there is another chemical potential which corresponds to the right-handed electrons (see also \cite{shiva3}).

The Abelian Chern-Simons term emerges in the effective action of the $U_Y(1)$ gauge fields, due to the chiral coupling of the Abelian hypercharge gauge fields to the fermions in the symmetric phase. The inclusion of the anomalous term in the magnetohydrodynamics (MHD) equations, results in the anomalous magnetohydrodynamics (AMHD) equations. This is an important term which leads to the well known chiral magnetic effect (CME). The AMHD equations describe the coupling between the hypercharge gauge fields, and the velocity and the number densities of the particles. Some authors have extensively studied the leptogenesis, the baryogenesis and the evolution of hypermagnetic fields in the context of the AMHD equations, but without considering the potential role of the velocity field  \cite{{26},{27},{28},{30},{31},{32},{33},{34},{shiva3}}.

The imbalanced chiral plasma which is affected by the CME in the presence of a magnetic field, is also influenced by the chiral vortical effect (CVE) when there is vorticity in the plasma \cite{{35},{36},{35-1},{Anand}}.
The chiral magnetic and vortical currents corresponding to these effects emerge in the AMHD equations and play important roles in the evolution of the cosmological magnetic fields and the matter-antimatter asymmetries. In the broken phase, the chiral magnetic and vortical currents have the form $\vec{J}_{\mathrm{cm}}\varpropto(\mu_{R}-\mu_{L})\vec{B}$, and $\vec{J}_{\mathrm{cv}}\varpropto\left(\mu_{R}^{2}-\mu_{L}^{2}\right) \vec{\omega}$, respectively, where $\vec{B}$ is the Maxwellian magnetic field, $\vec{\omega}=\vec{\nabla}\times \vec{v}$ is the vorticity, and $\mu_{R}$ and $\mu_{L}$ are the right-handed and the left-handed chemical potentials of the particles \cite{{35}}.\footnote{The forms of these currents are different in the symmetric phase, and will be presented later.} These currents are the macroscopic manifestation of the triangle anomaly in the chiral theory \cite{cveis,{dts1},{dts2},{dts3},{shi},{Omer}}. 
The notable fact about these currents is that they have a topological origin and are non-dissipative; therefore, they do not contribute to the entropy production \cite{kharseev}. 

The effects of the chiral magnetic and vortical currents on the evolution of the large scale magnetic fields and the matter-antimatter asymmetries have been investigated by the authors of Ref.\ \cite{36}. They have considered an incompressible fluid that has a fully non-helical vorticity field in the same direction as the magnetic field, but have ignored the viscosity damping effect. Recently, the effect of chiral anomaly on the evolution of the magnetohydrodynamic turbulence has been studied, as well \cite{37}. However, the effects of the chiral vorticity and the viscosity have not been taken into account. In another work related to the evolution of the magnetic fields in the neutron stars, the authors have taken the chiral magnetic effect into account, while considering the axial number density as a time and space dependent variable \cite{Dvornikov}. Then, they have added a new pseudoscalar term $\vec{\nabla}.S(t,x)$ to the evolution equation of $n_{5}(t,x)$, where $S(t,x)$ is the mean spin in the magnetized plasma. They have shown that the new term $\vec{\nabla}.S(t,x)$ produces chirality $\mu_{5}(t,x)$, and as a result, the chiral magnetic effect leads to the amplification of the seed magnetic field.

The main purpose of this paper is to present a simple model which starts with an initial chiral vorticity and describes not only the generation of the hypermagnetic field due to the CVE, but also its subsequent evolution which is mainly due to the CME, before the electroweak phase transition. Indeed, we present the correct form of the chiral vortical coefficient in the symmetric phase and show that, unlike the previous studies, the hypermagnetic field can be produced from zero initial value in the presence of the chiral vorticity. Furthermore, we show that the hypermagnetic field can be strengthened due to the CME. We also investigate the effects of these chiral vortical and magnetic currents on the evolution of the matter-antimatter asymmetries. 
In our model we use fully helical monochromatic hypermagnetic and vorticity fields. Since the hypermagnetic field is fully helical, $\vec{\nabla} \times \vec{B}\varpropto \vec{B}$, it has no influence on the evolution of the velocity or the vorticity fields \cite{38}. We also investigate the effects of the viscosity on the evolution of the vorticity and the hypermagnetic fields, and therefore, on the matter-antimatter asymmetries \cite{banerjee}. We also present the correct form of the fluid helicity in the symmetric phase which has been written incorrectly in some of the previous studies.

This paper is organized as follows. In Sec.\ \ref{x1}, we present the anomalous magnetohydrodynamics equations, and obtain the vorticity and the helicity coefficients in terms of the fermionic chemical potentials in the symmetric phase. In Sec.\ \ref{x3}, we consider the Abelian anomalous effects and derive the dynamical evolution equations of the fermionic asymmetries. In Sec.\ \ref{x4}, we solve the set of coupled differential equations numerically for the hypermagnetic field, the vorticity field, and the baryon and the first-generation lepton asymmetries. Finally, in Sec.\ \ref{x5}, we present our results and conclude.

 \section{ANOMALOUS MAGNETOHYDRODYNAMICS}\label{x1}
 
In this section, we briefly review the AMHD equations in the expanding Universe. Magnetohydrodynamics is the study of the electrically conducting fluids, combining both the principles of the fluid dynamics and the electromagnetism. 
In the imbalanced chiral plasma, the magnetic field and the vorticity induce the chiral magnetic effect (CME) and the chiral vortical effect (CVE), respectively. The CME is the generation of the electric current parallel to an external magnetic field, whereas the CVE is the generation of the electric current along the vorticity field.
In the presence of the Abelian anomaly, the MHD equations are generalized to the AMHD equations.
The evolution equations of the neutral plasma in the expanding Universe are given as (see Refs.\ \cite{{6},{dettmann},{Subramanian}} and also Appendix $\bf{A}$ and $\bf{B}$ for details)

\begin{equation}\label{eq1}
\frac{1}{R}\vec{\nabla} .\vec{E}_{Y}=0,\qquad\qquad\qquad\qquad\frac{1}{R}\vec{\nabla}.\vec{B}_{Y}=0,
\end{equation}

\begin{equation}\label{eq2}
\frac{\partial \vec{B}_{Y}}{\partial t}+2H\vec{B}_{Y}=-\frac{1}{R}\vec{\nabla}\times\vec{ E}_{Y},
\end{equation}	
\begin{equation}\label{eq2.1}\vec{J}_{\mathrm{Ohm}}=\sigma\left(\vec{E}_{Y}+\vec{v}\times\vec{B}_{Y}\right),
\end{equation}
\begin{equation}\label{eq3}
\vec{J}=\vec{J}_{\mathrm{Ohm}}+\vec{J}_{\mathrm{cv}}+\vec{J}_{\mathrm{cm}}=\frac{1}{R}\vec{\nabla}\times\vec{B}_{Y}-\left(\frac{\partial \vec{E}_{Y}}{\partial t}+2H\vec{E}_{Y}\right),
\end{equation}
\begin{equation}\label{eq3.1}
\vec{J}_{\mathrm{cv}}=c_{\mathrm{v}}\vec{\omega},
\end{equation}

\begin{equation}\label{eq3.2}
\vec{J}_{\mathrm{cm}}=c_{B}\vec{B}_{Y},
\end{equation}

\begin{equation}\label{eq4}
\begin{split}
\left[\frac{\partial}{\partial t}+\frac{1}{R}(\vec{v}.\vec{\nabla})+H\right]\vec{v}+\frac{\vec{v}}{\rho+p}\frac{\partial p}{\partial t}=\\-\frac{1}{R}\frac{\vec{\nabla} p}{\rho+p}+\frac{\vec{J}\times\vec{B}_{Y}}{\rho+p}+\frac{\nu}{{R}^{2}}\left[\nabla^{2}\vec{v}+\frac{1}{3}\vec{\nabla}(\vec{\nabla}.\vec{v})\right],
\end{split}
\end{equation}
\begin{equation}\label{eq5}
\vec{\omega}=\frac{1}{R}\vec{\nabla}\times\vec{v},
\end{equation}

\begin{equation}\label{eq7}
\frac{\partial \rho}{\partial t}+\frac{1}{R}\vec{\nabla}.\left[(\rho+p)\vec{v}\right]+3H(\rho+p)=0,
\end{equation}	
where, $\rho$ and $p$ are the energy density and the pressure of the fluid,  $\sigma$ is the electrical conductivity, $R$ is the scale factor, $H=\dot{R}/R$ is the Hubble parameter, and $\nu$ is the kinematic viscosity. 
Furthermore, $\vec{v}$ and $\vec{\omega}$ are the bulk velocity and the vorticity of the plasma, and the currents $\vec{J}_{\mathrm{Ohm}}$, $\vec{J}_{\mathrm{cv}}$, and $\vec{J}_{\mathrm{cm}}$ are the Ohmic current, the chiral vortical current, and the chiral magnetic current, respectively.\footnote{In some work in the literature, only the right-handed currents are considered, resulting in the following simplification: $\vec{J}_{\mathrm{cv}} = \vec{J}_{\mathrm{cv,R}} =\vec{J}_{\mathrm{cv}}^5$ and $\vec{J}_{\mathrm{cm}} = \vec{J}_{\mathrm{cm,R}} =\vec{J}_{\mathrm{cm}}^5$, see for example \cite{Anand}.}
The latter is the one that promotes the ordinary MHD equations to the AMHD equations. Note also that the terms like $2H\vec{B}_{Y}$ and $2H\vec{E}_{Y}$ in Eqs.\ (\ref{eq2}) and (\ref{eq3}) which contain the Hubble parameter $H$, are due to the expansion of the Universe. The vorticity and the helicity coefficients $c_{\mathrm{v}}$ and $c_{B}$ appearing in Eqs.\ (\ref{eq3.1}) and (\ref{eq3.2}) are as follows \cite{s.long, 31, 40} 

\begin{equation}\label{eq22}
c_{\mathrm{v}}(t)=\frac{{g'}}{16\pi^{2}}\sum_{i=1}^{n_{G}}\Big(-Y_{R}\mu_{R_{i}}^{2}+Y_{L}\mu_{L_{i}}^{2}N_{w}-Y_{d_{R}}\mu_{d_{R_{i}}}^{2}N_{c}-Y_{u_{R}}\mu_{u_{R_{i}}}^{2}N_{c}+Y_{Q}\mu_{Q_{i}}^{2}N_{c}N_{w}\Big),	
\end{equation}

\begin{equation}\label{eq25}
\begin{split}
c_{B}(t)=&
-\frac{g'^{2}}{8\pi^{2}}\sum_{i=1}^{n_{G}}\Big[-\left(\frac{1}{2}\right)Y_{R}^{2}\mu_{R_{i}}-\left(\frac{-1}{2}\right)Y_{L}^{2}\mu_{L_{i}}N_{w}-\left(\frac{1}{2}\right)Y_{d_{R}}^{2}\mu_{d_{R_{i}}}N_{c}\\&-\left(\frac{1}{2}\right)Y_{u_{R}}^{2}\mu_{u_{R_{i}}}N_{c}-\left(\frac{-1}{2}\right)Y_{Q}^{2}\mu_{Q_{i}}N_{c}N_{w}\Big],	
\end{split}
\end{equation}
%
%
%
where, $n_{G}$ is the number of generations, and $N_{c}=3$ and $N_{w}=2$ are the ranks of the non-Abelian SU$(3)$ and SU$(2)$ gauge groups, respectively. Furthermore, $\mu_{L_i}$($\mu_{R_i}$), $\mu_{Q_i}$, and $\mu_{{u_R}_i}$ ($\mu_{{d_R}_i}$) are the common chemical potentials of left-handed (right-handed) leptons, the left-handed quarks with different colors, and up (down) right-handed quarks with different colors, respectively. Moreover, \lq{\textit{i}}\rq\ is the generation index, and the corresponding hypercharges are 
\begin{equation}\label{eq23}
\begin{split}
&Y_{L}=-1,\qquad\qquad Y_{R}=-2,\\
&Y_{u_{R}}=\frac{4}{3},\qquad\qquad Y_{d_{R}}=-\frac{2}{3},\qquad\qquad Y_{Q}=\frac{1}{3}.
\end{split}
\end{equation}
Substituting the above constants in Eqs.\ (\ref{eq22}) and (\ref{eq23}) results in 
\begin{equation}\label{eq24}
c_{\mathrm{v}}(t)=\frac{{g'}}{16\pi^{2}}\sum_{i=1}^{n_{G}}\left(2\mu_{R_{i}}^{2}-2\mu_{L_{i}}^{2}+2\mu_{d_{R_{i}}}^{2}-4\mu_{u_{R_{i}}}^{2}+2\mu_{Q_{i}}^{2}\right),	
\end{equation}



\begin{equation}\label{eq26}
c_{B}(t)=-\frac{g'^{2}}{8\pi^{2}}\sum_{i=1}^{n_{G}}\left(-2\mu_{R_{i}}+\mu_{L_{i}}-\frac{2}{3}\mu_{d_{R_{i}}}-\frac{8}{3}\mu_{u_{R_{i}}}+\frac{1}{3}\mu_{Q_{i}}\right).	
\end{equation}


In a previous study \cite{34}, we took the CME into account but neglected the CVE. There, we made some assumptions and simplified the helicity coefficient $c_{B}$ accordingly. In this paper, we take both the CME and the CVE into account.
In the following, we make the same assumptions and simplify the vorticity coefficient $c_{\mathrm{v}}$, as well. 

We assume that all quark Yukawa interactions\footnote{They are: up-type Yukawa in processes $u_R^i\bar{d}_L^i\leftrightarrow\phi^{(+)}$ and $u_R^i\bar{u}_L^i\leftrightarrow\phi^{(0)}$, down-type Yukawa in processes $d_R^j\bar{u}_L^i \leftrightarrow \phi^{(-)}$ and $d_R^j\bar{d}_L^i \leftrightarrow \tilde{\phi}^{(0)}$, and their conjugate reactions \cite{s.long}.} are in equilibrium. Moreover, because of the flavor mixing in the quark sector, we assume that all up or down quarks which belong to different generations with distinct handedness have the same chemical potential. Then, the following equilibrium conditions are obtained \cite{34,17}.

\begin{equation}\label{eq27}
\mu_{u_{R}}-\mu_{Q}=\mu_{0},
\end{equation}
\begin{equation}\label{eq28}
\mu_{d_{R}}-\mu_{Q}=-\mu_{0}.
\end{equation}
Where, $\mu_{0}$, $\mu_{Q}$, and $\mu_{u_{R}}$ ($\mu_{d_{R}}$) are the chemical potentials of the Higgs field, the left-handed up or down quarks, and the right-handed up (down) quarks, respectively. Then, for simplicity, we assume that the Higgs asymmetry is zero and obtain \cite{27,34}
\begin{equation}\label{eq29}
\mu_{u_{R}}=\mu_{d_{R}}=\mu_{Q}.
\end{equation}
Using Eq.\ (\ref{eq29}), we simplify Eqs.\ (\ref{eq24}) and (\ref{eq26}), and obtain 
\begin{equation}\label{eq30}
c_{\mathrm{v}}(t)=\frac{{g'}}{8\pi^{2}}\sum_{i=1}^{n_{G}}\left(\mu_{R_{i}}^{2}-\mu_{L_{i}}^{2}\right)	
\end{equation}
and
\begin{equation}\label{eq31}
c_{B}(t)=-\frac{g'^{2}}{8\pi^{2}}\sum_{i=1}^{n_{G}}\left(-2\mu_{R_{i}}+\mu_{L_{i}}-3\mu_{Q}\right).	
\end{equation}
We assume that only the contributions of the baryonic and the first-generation leptonic chemical potentials to the helicity and the vorticity coefficients are non-negligible. 
Then, Eqs.\ (\ref{eq30}) and (\ref{eq31}) reduce to the forms
\begin{equation}\label{eq32}
c_{\mathrm{v}}(t)=\frac{{g'}}{8\pi^{2}}\left(\mu_{e_{R}}^{2}-\mu_{e_{L}}^{2}\right),	
\end{equation}
\begin{equation}\label{eq33}
c_{B}(t)=-\frac{g'^{2}}{8\pi^{2}}\left(-2\mu_{e_{R}}+\mu_{e_{L}}-\frac{3}{4}\mu_{B}\right),	
\end{equation}
where we have also used the equation $\mu_{Q}=\frac{1}{12}\mu_{B}$ \cite{34}. The time-dependent coefficients $c_{\mathrm{v}}(t)$ and $c_{B}(t)$ evolve in accordance to the evolution of their constituents, the evolution equations of which will be obtained in the next section.
Like what is usually done in the ordinary MHD equations, the displacement current will be neglected in the following. It should be noted that, neglecting the displacement current in the comoving frame is equivalent to neglecting the term $\partial_{t} \vec{E}_{Y}+2H\vec{E}_{Y}$ in the Lab frame. Using Eqs.\ (\ref{eq2.1}) and (\ref{eq3}) with the mentioned assumption, the hyperelectric field can be obtained as

\begin{equation}\label{eq8}
\vec{E}_{Y}=\frac{1}{\sigma R}\vec{\nabla}\times\vec{B}_{Y}-\frac{c_{\mathrm{v}}}{\sigma}\vec{\omega}-\frac{c_{B}}{\sigma}\vec{B}_{Y}-\vec{v}\times\vec{B}_{Y}.
\end{equation}
Putting the above expression for the hyperelectric field into Eq.\ (\ref{eq2}), the evolution equation of the hypermagnetic field can be obtained as
\begin{equation}\label{eq9}
\frac{\partial\vec{ B}_{Y}}{\partial t}+\frac{\vec{ B}_{Y}}{t}=\frac{1}{\sigma R^{2}}\nabla^{2}\vec{B}_{Y}+\frac{c_{\mathrm{v}}}{\sigma R}\vec{\nabla}\times\vec{\omega}+\frac{c_{B}}{\sigma R}\vec{\nabla}\times\vec{B}_{Y}+\frac{1}{R}\vec{\nabla}\times(\vec{v}\times\vec{B}_{Y}),
\end{equation}
where we have used the equation $H=1/2t$ for the radiation dominated era. 

Since $ \vec{\nabla}.\vec{B}_{Y}=0$, the hypermagnetic field can be written as $\vec{B}_{Y}=(1/R)\vec{\nabla}\times\vec{A}_{Y}$, where $\vec{A}_{Y}$ is the vector potential of the hypermagnetic field. Let us consider an incompressible fluid in the comoving frame \cite{36,37}, which leads to the condition of $\partial_{t}\rho+3H(\rho+p)=0$ in the lab frame. Then, combining this condition with the continuity equation (\ref{eq7}) results in $\vec{\nabla} .\vec{v}=0$.
Therefore, in analogy with the hypermagnetic field, the velocity field can be written as $\vec{v}=(1/R)\vec{\nabla} \times\vec{S}$, where $\vec{S}$ is the vector potential of the velocity field.
In this work, we concentrate on the fully helical hypermagnetic and vorticity fields; To have such fields, we choose the same non-trivial Chern-Simons wave configuration\footnote{As mentioned in the Introduction, the coherent magnetic fields in the intergalactic medium have been inferred to be helical \cite{Chen2}. Therefore, we have chosen a helical configuration for the hypermagnetic field \cite{69,70}. This topologically non-trivial configuration, with Chern-Simons number density $n_{CS}\varpropto k^{\prime}\gamma^{2}(t)$, has been used extensively to solve the magnetohydrodynamic (MHD) equations \cite{27,28,31,34,shiva3,73}. Furthermore, it has been introduced as an exact single-mode solution to the chiral MHD equations \cite{74,75}. Moreover, the four fully helical configurations are $\vec{A}_{Y}=\gamma(t)( \sin kz ,\cos kz, 0)$, $\vec{A}_{Y}=\gamma(t)(\cos kz,- \sin kz , 0)$, $\vec{A}_{Y}=\gamma(t)( \cos kz,\sin kz , 0)$, and $\vec{A}_{Y}=\gamma(t)( -\sin kz ,\cos kz, 0)$. The first two (last two) have positive (negative) helicity, and their unit vectors, along with $\hat{z}$, form orthonormal bases \cite{76}.}\cite{69,70} for both of their vector potentials, in order to have maximum efficacy. That is, 
\begin{equation}\label{eq10}
\vec{A}_{Y}=\gamma(t)(\sin kz , \cos kz, 0),
\end{equation}
and
\begin{equation}\label{eq11}
\vec{S}=r(t)(\sin kz , \cos kz, 0),
\end{equation}
where $\gamma (t)$ and $r(t)$ are the time-dependent amplitudes of the vector potentials $\vec{A}_{Y}$ and $\vec{S}$, respectively. Using these configurations, we get $\vec{B}_{Y}=(1/R)k \vec{A}_{Y}$, $\vec{v}=(1/R)k \vec{S}$, and $\vec{\omega}=(1/R)k \vec{v}$ for the hypermagnetic, the velocity and the vorticity fields. In the following, $\vec{\omega}$ will be replaced by $(1/R)k \vec{v}$, wherever appropriate.  

Let us compute the ensemble average of the hypermagnetic field energy density by using the aforementioned simple configuration as
 \begin{equation}\label{eq12}
 \begin{split}
 E_{B}(t)&=\frac{1}{2}\langle \vec{B}_{Y}(x,t) . \vec{B}_{Y}(x,t) \rangle\\&=\frac{1}{2}B_{Y}^{2}(t)=\frac{1}{2R^{2}}k^{2}\gamma^{2}(t),
 \end{split}
 \end{equation}
where the angle brackets denote the ensemble average. Similarly, the hypermagnetic helicity density can be computed as
 \begin{equation}\label{eq13}
 H_{B}(t)=\langle \vec{A}_{Y}(x,t) . \vec{B}_{Y}(x,t) \rangle=\frac{k}{R}\gamma^{2}(t).
 \end{equation}
It can be seen that $ E_{B}(t)=(k/2R)H_{B}(t)$, which indicates that the hypermagnetic field is fully helical.
 
In analogy with the hypermagnetic field, the fluid kinetic energy and the fluid helicity can be defined as
 \begin{equation}\label{eq14}
 	\begin{split}
 E_{\mathrm{v}}(t)&=\frac{\rho}{2}\langle \vec{v}.\vec{v} \rangle =\frac{\rho}{2}v^{2}(t),
 \end{split}
 \end{equation}
 and 
 
\begin{equation}\label{eq15}
\begin{split}
H_{\mathrm{v}}(t)&=\sum_{i=1}^{n_{G}}\Big[\Big(\frac{1}{24}\Big)\Big(T_{R_{i}}^{2}+T_{L_{i}}^{2}N_{w}+T_{d_{R_{i}}}^{2}N_{c}+T_{u_{R_{i}}}^{2}N_{c}+T_{Q_{i}}^{2}N_{c}N_{w}\Big)\\&+\Big(\frac{1}{8\pi^{2}}\Big)\Big(\mu_{R_{i}}^{2}+\mu_{L_{i}}^{2}N_{w}+\mu_{d_{R_{i}}}^{2}N_{c}+\mu_{u_{R_{i}}}^{2}N_{c}+\mu_{Q_{i}}^{2}N_{c}N_{w}\Big)\Big]\langle \vec{v}.\vec{w} \rangle\\&=\sum_{i=1}^{n_{G}}\left[\frac{15}{24}T^{2}+\Big(\frac{1}{8\pi^{2}}\Big)\Big(\mu_{R_{i}}^{2}+2\mu_{L_{i}}^{2}+12\mu_{Q}^{2}\Big)\right]\frac{k}{R}v^{2}(t),
\end{split}
\end{equation}
respectively \cite{{cveis},{Sadofyev},{Avkhadiev},{Sadofyev22}}. In Eq.\ (\ref{eq15}), we have assumed that all particles are in thermal equilibrium, and as mentioned earlier, $\mu_{d_{R}}=\mu_{u_{R}}=\mu_{Q}$. 
It can be seen that the time-dependent temperatures and chemical potentials play important roles in the fluid helicity, and even with constant velocity, the fluid helicity decreases as $R^{-3}$ due to the expansion of the Universe. 



Using the simple configurations for the vector potentials of the hypermagnetic and the velocity fields as given by Eqs.\ (\ref{eq10}) and (\ref{eq11}), and their consequent relations, $\vec{B}_{Y}=(1/R)\vec{\nabla}\times\vec{A}_{Y}=(1/R)k \vec{A}_{Y}$ and $\vec{\omega}=(1/R)\vec{\nabla}\times\vec{v}=(1/R)k \vec{v}$, Eqs.\ (\ref{eq8}) and (\ref{eq9}) reduce to the forms 

\begin{equation}\label{eq16}
\vec{E}_{Y}=\frac{k^{\prime}}{\sigma }\vec{B}_{Y}-\frac{c_{\mathrm{v}}}{\sigma }k^{\prime}\vec{v}-\frac{c_{B}}{\sigma}\vec{B}_{Y},
\end{equation}
and
\begin{equation}\label{eq17}
\frac{\partial {\vec{B}}_{Y}(t)}{\partial t}+\frac{ {\vec{B}}_{Y}(t)}{ t}=\frac{-{k^{\prime}}^{2}}{\sigma }{\vec{B}}_{Y}(t)+\frac{c_{\mathrm{v}}}{\sigma}{k^{\prime}}^{2}{\vec{v}}(t)+\frac{c_{B}}{\sigma }k^{\prime} {\vec{B}}_{Y}(t),
\end{equation}
respectively, where $k^{\prime}=k/R=kT$. It can be seen that the length scale of the hypermagnetic field increases due to the expansion of the Universe.
Note that both the hypermagnetic and the velocity fields are in the same direction; thus, the advection term $\vec{v}\times\vec{B}_{Y}$ in Eqs.\ (\ref{eq8}) and (\ref{eq9}) has been set to zero.

Let us now consider the evolution equation of the velocity field. Neglecting the displacement current in Eq.\ (\ref{eq3}), the total current becomes $\vec{J}=(1/R)\vec{\nabla}\times\vec{B}_{Y}$; therefore, $\vec{J}\times\vec{B}_{Y}$ vanishes in Eq.\ (\ref{eq4}).
Furthermore, the incompressibility condition of the fluid, $\partial_{t}\rho+3H(\rho+p)=0$, not only leads to $\vec{\nabla} .\vec{v}=0$, as stated earlier, but also ensures that $H\vec{v}+\vec{v}\partial_{t}p/(\rho+p)=0$ in Eq.\ (\ref{eq4}). After neglecting the gradient terms in Eq. (\ref{eq4}),\footnote{The term $(\vec{v}.\vec{\nabla})\vec{v}$ is neglected since it is next to leading order. Furthermore, $\vec{\nabla} p=0$ because the fluid pressure is only time-dependent.} the evolution equation of the velocity field simplifies to
 
\begin{equation}\label{eq19}
\frac{\partial \vec{v}}{\partial t}=-\nu{k^{\prime}}^{2}\vec{v}.
\end{equation}
Note that in the radiation dominated era, only the shear viscosity contributes to the non-ideal stress energy tensor and the bulk viscosity becomes zero. In the next section, the evolution equations of the fermion numbers will be obtained.

\section{ABELIAN ANOMALY AND FERMION NUMBER VIOLATION}\label{x3}

In this section, we briefly review the $U_Y(1)$ Abelian anomaly equations, and obtain the evolution equations of the leptonic and the baryonic asymmetries in the symmetric phase. Before the electroweak phase transition, in contrast to the broken phase, the fermion numbers are violated, due to the fact that the coupling of the hypercharge fields to the fermions is chiral. This shows up in the $U_Y(1)$ Abelian anomaly equations\cite{{26}}. These anomaly equations for the first-generation leptons are
  
\begin{equation}
\begin{split}
&\partial_{\mu} j_{{e}_R}^{\mu}=-\frac{1}{4}\left(Y_{R}^{2}\right)\frac{g'^{2}}{16 \pi^2}Y_{\mu\nu}\tilde{Y}^{\mu\nu}=\frac{g'^{2}}{4\pi^{2}}\vec{E}_{Y}.\vec{B}_{Y},\\
&\partial_{\mu} j_{{e}_L}^{\mu}=\partial_{\mu} j_{{\nu}_{e}^{L}}^{\mu}=\frac{1}{4}\left(Y_{L}^{2}\right)\frac{g'^{2}}{16\pi^2}Y_{\mu\nu}\tilde{Y}^{\mu\nu}=-\frac{g'^{2}}{16\pi^{2}}\vec{E}_{Y}.\vec{B}_{Y}.
 \end{split}
\end{equation}
In addition to the Abelian anomaly that violates the lepton numbers, the perturbative chirality flip reactions should also be considered in the evolution equations of the leptonic asymmetries as

\begin{equation}
\begin{split}
&\frac{d\eta_{{e}_{R}}}{dt}=\frac{g'^{2}}{4\pi^{2} s}\langle\vec{E}_{Y}.\vec{B}_{Y}\rangle+2\Gamma_{RL}\left(\eta_{e_{L}}-\eta_{e_{R}}\right),\\
&\frac{d\eta_{{e}_{L}}}{dt}=\frac{d\eta_{{\nu}_{e}^{L}}}{dt}=-\frac{g'^{2}}{16\pi^{2} s}\langle\vec{E}_{Y}.\vec{B}_{Y}\rangle+\Gamma_{RL}\left(\eta_{e_{R}}-\eta_{e_{L}}\right),
 \end{split}
\end{equation} 
where, $\eta_{f}=(n_{f}-n_{\bar{f}})/s$ with $f=e_{R},e_{L},\nu_{e}^{L}$ is the fermion asymmetry, $s=2\pi^{2}g^{*}T^{3}/45$ is the entropy density, and $g^{*}=106.75$ is the effective number of relativistic degrees of freedom. It should be noted that we are assuming  $\eta_{{e}_{L}}\approx \eta_{{\nu}_{e}^{L}}$, based on the fast SU(2) interactions in the SU(2) doublet. The chirality flip rate $\Gamma_{RL}$ that appears in the above equations is \cite{28}   

\begin{equation}
\Gamma_{RL}=5.3\times10^{-3}h_{e}^{2}(\frac{m_{0}}{T})^{2}T=\left(\frac{\Gamma_{0}}{2t_{EW}}\right)\left(\frac{1-x}{\sqrt{x}}\right),
\end{equation}   
where the variable $x=\frac{t}{t_{EW}}=(\frac{T_{EW}}{T})^{2}$, in accordance with the Friedmann law, $t_{EW}=\frac{M_{0}}{2T_{EW}^{2}}$,  $M_{0}=M_{Pl}/1.66\sqrt{g^{*}}$, and $M_{Pl}$ is the Plank mass. In addition, $h_{e}=2.94\times10^{-6}$ is the Yukawa coupling of the right-handed electrons, $\Gamma_{0}=121$, and $m_{0}^{2}(T)=2D T^{2}(1-T_{EW}^{2}/T^{2})$ is the temperature-dependent effective Higgs mass at zero momentum and zero Higgs vacuum expectation value. The coefficient $2D\sim 0.377$ in the expression for $ m_{0}^{2}(T)$ has contributions coming from the known masses of gauge bosons $m_{Z}$ and $m_{W}$, the top quark mass $m_{t}$, and the zero-temperature Higgs mass\cite{28}. Using the expression for the fermionic chemical potential, $\mu_{f}=6(n_{f}-n_{\bar{f}})/T^{2}$, and the changes of variables, $\xi_{f}=\mu_{f}/T$ and $\eta_{f}=\xi_{f}T^{3}/6s$, we obtain
  
 \begin{equation}\label{eqxi}
 \begin{split}
  &\frac{d\xi_{{e}_{R}}}{dt}=\frac{3 g'^{2}}{2\pi^{2} T^{3}}\langle\vec{E}_{Y}.\vec{B}_{Y}\rangle+2\Gamma_{RL}\left(\xi_{e_{L}}-\xi_{e_{R}}\right),\\
  &\frac{d\xi_{{e}_{L}}}{dt}=\frac{d\xi_{{\nu}_{e}^{L}}}{dt}=-\frac{3 g'^{2}}{8\pi^{2} T^{3}}\langle\vec{E}_{Y}.\vec{B}_{Y}\rangle+\Gamma_{RL}\left(\xi_{e_{R}}-\xi_{e_{L}}\right).
  \end{split}
  \end{equation}
By considering the conservation law $\eta_{B}/3 -\eta_{L_{i}}= \mbox{const.}$ and the evolution equations of the lepton asymmetries, the evolution equation of the baryon asymmetry can be obtained as
  \begin{equation}\label{eq39}
  \frac{1}{3}\frac{d\xi_{B}}{dt}=\frac{d\xi_{e_{R}}}{dt}+2\frac{d\xi_{e_{L}}}{dt}=\frac{3 g'^{2}}{4\pi^{2} T^{3}}\langle\vec{E}_{Y}.\vec{B}_{Y}\rangle.
  \end{equation}
In the above equations, we need to know the exact form of $\langle\vec{E}_{Y}.\vec{B}_{Y}\rangle$. We use Eq.\ (\ref{eq16}) and obtain
 \begin{equation}\label{eq42}
 \langle\vec{E}_{Y}.\vec{B}_{Y}\rangle=\frac{k^{\prime}}{\sigma} B_{Y}^{2}(t)-\frac{c_{B}}{\sigma} B_{Y}^{2}(t)-\frac{c_{\mathrm{v}}k^{\prime}}{\sigma} \langle\vec{v}(t).\vec{B}_{Y}(t)\rangle ,
 \end{equation}
where the vorticity coefficient, $c_{\mathrm{v}}$, and the helicity coefficient, $c_{B}$, are given by Eqs.\ (\ref{eq32},\ref{eq33}), respectively.
Then, using $\sigma=100T$, $R=1/T$, $\nu\simeq1/(5\alpha_{Y}^{2}T)$ \cite{{41},{banerjee}}, where $\alpha_{Y}=g'^{2}/4\pi$ is the fine-structure constant for the $U_Y(1)$ gauge fields, and the aforementioned expressions for $c_{\mathrm{v}}$ and $c_{B}$, Eqs.\ (\ref{eq42}), (\ref{eq17}), and (\ref{eq19}) become

\begin{equation}\label{eq44}
\begin{split}
\langle\vec{E}_{Y}.\vec{B}_{Y}\rangle=&\frac{B_{Y}^{2}(t)}{100} \left[\frac{k^{\prime}}{T}-\frac{g'^{2}}{4\pi^{2}}\left(\xi_{e_{R}}-\frac{\xi_{e_{L}}}{2}+\frac{3}{8}\xi_{B}
\right)\right]\\&-\frac{g'}{800\pi^{2}}\left(\xi_{e_{R}}^{2}-\xi_{e_{L}}^{2}\right)k^{\prime}T \langle\vec{v}(t).\vec{B}_{Y}(t)\rangle,
\end{split}
\end{equation}
\begin{equation}\label{eq45}
\begin{split}
\frac{d B_{Y}(t)}{dt}=&\frac{B_{Y}(t)}{100}\left[-\frac{{k^{\prime}}^{2}}{T}+\frac{k^{\prime}g'^{2}}{4\pi^{2}}\left(\xi_{e_{R}}-\frac{\xi_{e_{L}}}{2}+\frac{3}{8}\xi_{B}\right)\right]\\&-\frac{ B_{Y}(t)}{ t}+\frac{g'}{800\pi^{2}}\left(\xi_{e_{R}}^{2}-\xi_{e_{L}}^{2}\right){k^{\prime}}^{2}T\langle\vec{v}(t).\hat{B}_{Y}(t)\rangle,
\end{split}
\end{equation}
\begin{equation}\label{eq46}
 \frac{dv(t)}{dt}=-\frac{{k^{\prime}}^{2}}{5\alpha_{Y}^{2}T}v(t).
\end{equation}
With the choice of the vector potentials in Eqs.\ (\ref{eq10},\ref{eq11}), $\langle\vec{v}(t).\vec{B}_{Y}(t)\rangle \rightarrow v(t)B_Y(t)$ and $\langle\vec{v}(t).\hat{B}_{Y}(t)\rangle \rightarrow v(t)$. Using Eq.\ (\ref{eq44}), and the relations $y_{R}=10^{4}\xi_{e_{R}}$, $y_{L}=10^{4}\xi_{e_{L}}$, $x=t/t_{EW}=(T_{EW}/T)^{2}$, and $1\mbox{Gauss}\simeq2\times10^{-20} \mbox{GeV}^{2}$, we can rewrite Eqs.\ (\ref{eqxi}), (\ref{eq45}), and (\ref{eq46}) in the forms
 \begin{equation}\label{eq47}
 \begin{split}
 \frac{dy_{R}}{dx}=&\left[C_{1}-C_{2}\left(y_{R}-\frac{y_{L}}{2}+\frac{3}{8}y_{B}\right)\right]\left(\frac{B_{Y}(x}{10^{20}G}\right)^{2}x^{3/2}\\&-C_{3}\left(y_{R}^{2}-y_{L}^{2}\right)v(x)\left(\frac{B_{Y}(x)}{10^{20}G}\right)\sqrt{x}-\Gamma_{0}\frac{1-x}{\sqrt{x}}(y_{R}-y_{L}),
 \end{split}
 \end{equation}
  
   \begin{equation}\label{eq48}
   \begin{split}
  \frac{dy_{L}}{dx}=&-\frac{1}{4}\left[C_{1}-C_{2}\left(y_{R}-\frac{y_{L}}{2}+\frac{3}{8}y_{B}\right)\right]\left(\frac{B_{Y}(x)}{10^{20}G}\right)^{2}x^{3/2}\\&+\frac{C_{3}}{4}\left(y_{R}^{2}-y_{L}^{2}\right)v(x)\left(\frac{B_{Y}(x)}{10^{20}G}\right)\sqrt{x}+\Gamma_{0}\frac{1-x}{2\sqrt{x}}(y_{R}-y_{L}),
  \end{split}
  \end{equation}
  
  \begin{equation}\label{eq49} 
  \begin{split}
  \frac{dB_{Y}}{dx}=&\frac{C_{4}}{\sqrt{x}}\left[-\left(\frac{k}{10^{-7}}\right) +\frac{10^{3}\alpha_{Y}}{\pi}\left(y_{R}-\frac{y_{L}}{2}+\frac{3}{8}y_{B}\right)\right]B_{Y}(x)-\frac{B_{Y}(x)}{x}\\&+C_{5}\left(y_{R}^{2}-y_{L}^{2}\right)\frac{v(x)}{x^{3/2}},
  \end{split}
 \end{equation}
 
\begin{equation}\label{eq50}
\frac{dv(x)}{dx}=-\frac{C_{6}}{\sqrt{x}}v(x),
\end{equation}
where
\begin{equation}\label{eq51}
\begin{split}
 & C_{1}=25.78\left(\frac{k}{10^{-7}}\right),\qquad C_{2}=77.79,  \qquad C_{3}=0.0534\left(\frac{k}{10^{-7}}\right)\frac{g'^{3}}{4\pi^{4}},\\&
C_{4}= 0.356\left(\frac{k}{10^{-7}}\right), \quad C_{5}=89\times 10^{13}\frac{g'}{4\pi^{2}}\left(\frac{k}{10^{-7}}\right)^{2},\quad C_{6}=\frac{7.12}{\alpha_{Y}^{2}} \left(\frac{k}{10^{-7}}\right)^{2}.
\end{split}
\end{equation}
 
Following steps analogous to those for the derivation of Eq.\ (\ref{eq39}), we obtain the evolution equation of the baryon asymmetry in the form

\begin{equation}\label{eq54}
\begin{split}
\frac{dy_{B}}{dx}=&\frac{3}{2}\left[C_{1}-C_{2}\left(y_{R}-\frac{y_{L}}{2}+\frac{3}{8}y_{B}\right)\right]\left(\frac{B_{Y}(x)}{10^{20}G}\right)^{2}x^{3/2}\\&-\frac{3\sqrt{x}}{2}C_{3}\left(y_{R}^{2}-y_{L}^{2}\right)v(x)\left(\frac{B_{Y}(x)}{10^{20}G}\right),
\end{split}
\end{equation}
where $y_{B}=4\times10^{4} \pi^{2}g^{*}\eta_{B}/15$. The terms containing $v(x)$ in Eqs.\ (\ref{eq47}), (\ref{eq48}), (\ref{eq49}), and (\ref{eq54}) are due to the presence of the chiral vorticity in the plasma. In the next section we will solve this set of coupled differential equations numerically and discuss the results.

\section{NUMERICAL SOLUTION}\label{x4}
 
\begin{figure}[!ht]
	\subfigure[]{\label{fig:figure:im1}
		\includegraphics[width=.45\textwidth]{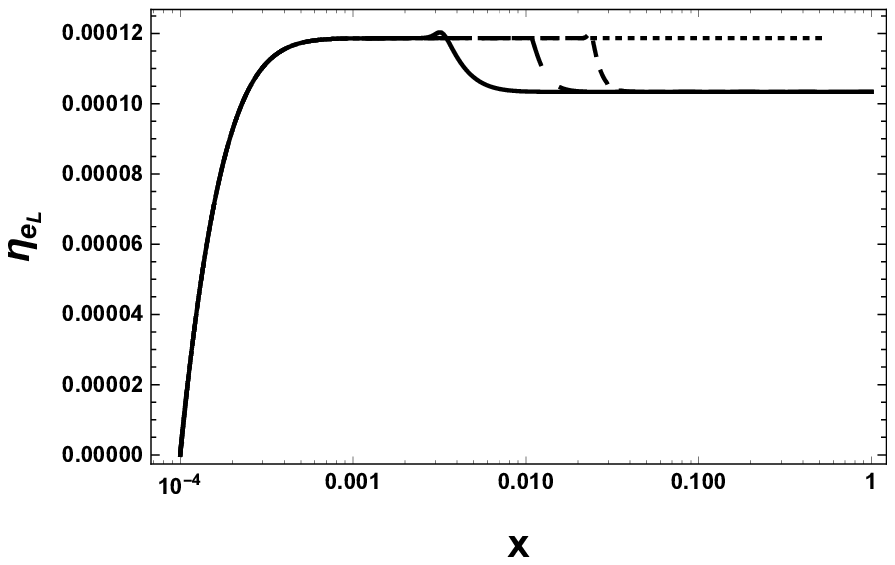}}
	\hspace{8mm}
	\subfigure[]{\label{fig:figure:im2} 
		\includegraphics[width=.45\textwidth]{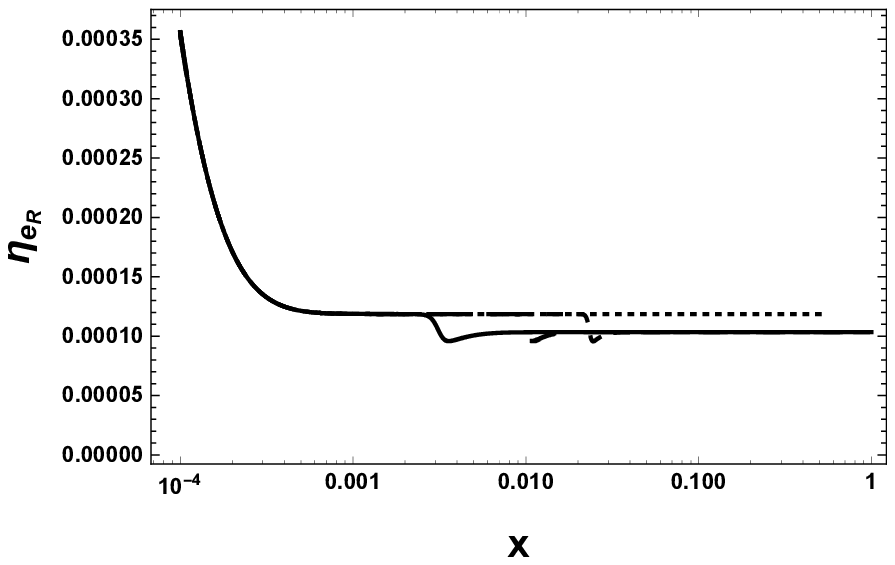}}
	\hspace{8mm}
	\subfigure[]{\label{fig:figure:im3} 
		\includegraphics[width=.45\textwidth]{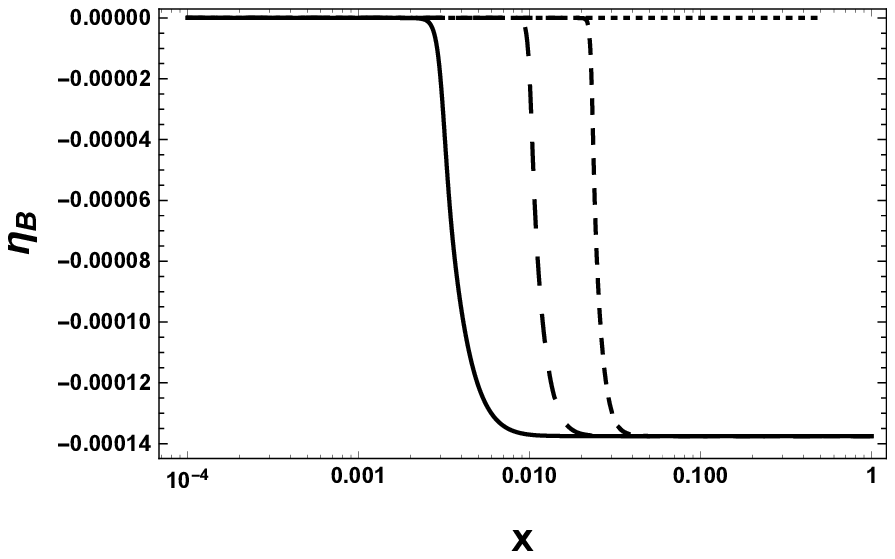}}
	\hspace{8mm}
	\subfigure[]{\label{fig:figure:in1}
		\includegraphics[width=.45\textwidth]{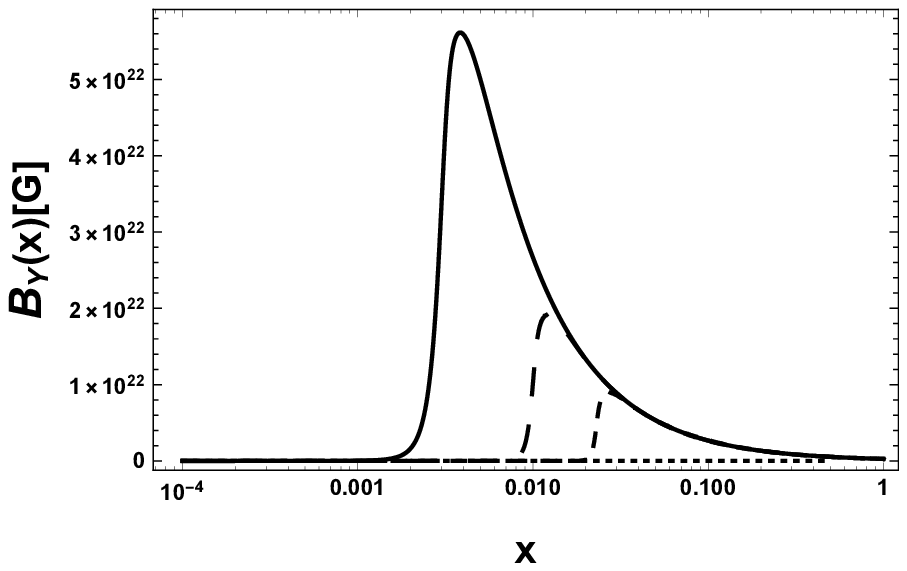}}
	\hspace{8mm}
	\subfigure[]{\label{fig:figure:in21}
	\includegraphics[width=.45\textwidth]{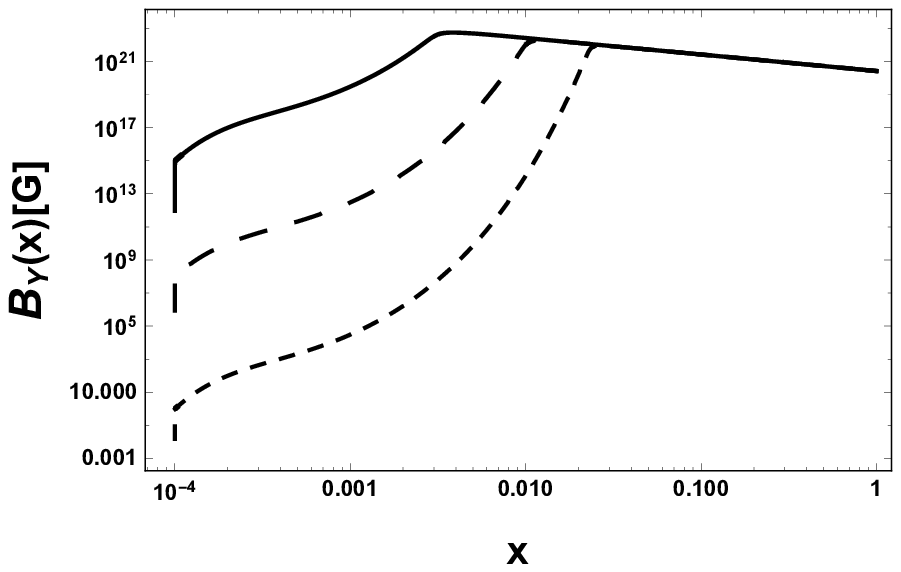}}
	\hspace{8mm}
	\subfigure[]{\label{fig:figure:in2} 
	\includegraphics[width=.45\textwidth]{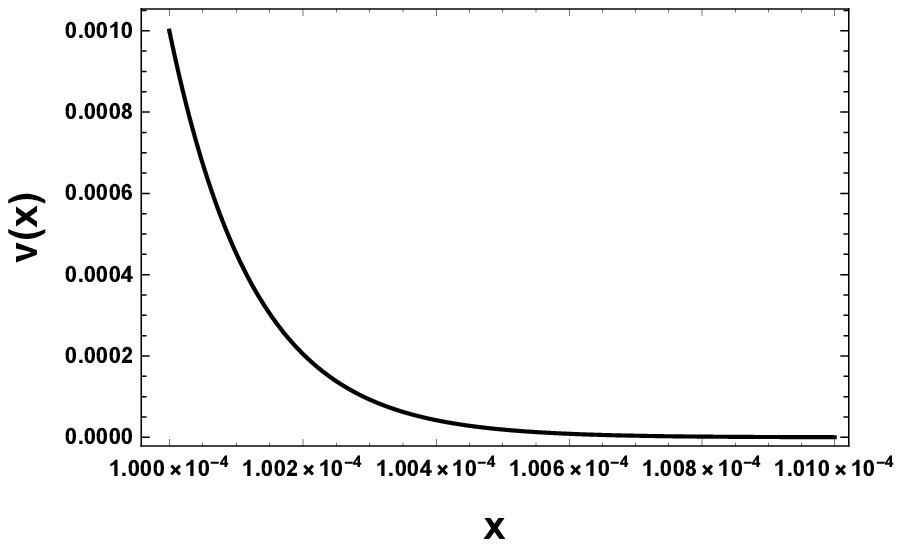}}
	
	\caption{\footnotesize Time plots of the lepton and the baryon asymmetries and the hypermagnetic field amplitude in the presence of the viscosity with the initial conditions $k=10^{-7}$, $B_{Y}^{(0)}=0$, $y_{R}^{(0)}=10^{3}$, and $y_{L}^{(0)}=y_{B}^{(0)}=0$. The solid line is for $v_{0}=10^{-3}$, large dashed line for $v_{0}=10^{-10}$, dashed line for $v_{0}=10^{-18}$, and dotted line for $v_{0}=0$.\\  a: Left-handed lepton asymmetry, $\eta_{e_{L}}$. \\ b: Right-handed lepton asymmetry, $\eta_{e_{R}}$.\\ c: Baryon asymmetry, $\eta_{B}$. \\ d: The hypermagnetic field amplitude, $B_{Y}$.\\ e: The log plot of  $B_{Y}$ (The case for $v_{0}=0$ yielding $B_{Y}=0$ cannot be displayed).\\ f: The velocity field amplitude for $v_{0}=10^{-3}$ and $10^{-4}\le x\le 1.01\times10^{-4}$.}		
	\label{fig1}
\end{figure}

In this section, we solve the set of coupled differential equations obtained in Sec.\ \ref{x3} numerically, and compare the results with the ones obtained in the non-vortical plasma. The equations are solved with the initial conditions $k=10^{-7}$, $B_{Y}^{(0)}=0$, $y_{R}^{(0)}=10^{3}$, $y_{L}^{(0)}=y_{B}^{(0)}=0$, and four different values for the initial velocity, $v^{(0)}=0, 10^{-18}, 10^{-10}$, and $10^{-3}$. The initial velocities are all within the domain of validity of the non-relativistic approximation. The results are shown in Fig.\ \ref{fig1}.

Figure \ref{fig1} shows that lepton asymmetries are equalized rather quickly by the chirality flip processes. As shown in the Figs.\ 1(a-d), if the initial velocity, and hence the vorticity, is zero, nothing else happens. That is, the lepton asymmetries remain constant, and the baryon asymmetry and the hypermagnetic field amplitude remain zero. However, if the initial vorticity is non-zero, the CVE causes $B_{Y}$ to grow extremely rapidly at the start of its evolution, essentially creating a seed field for it. By increasing the initial velocity, the seed field becomes stronger, and its ensuing growth due to the CME leads to yet larger values (see Fig.\ 1e). The maximum scale of this initial growth can be seen in Fig.\ 1f, which shows how quickly the initial velocity is damped by the viscosity. 

When $B_{Y}$ is produced, it grows until it reaches a maximum or saturation value at a critical time, and a concurrent transition occurs: the lepton and baryon asymmetries decrease rapidly (see Figs.\ 1(a-e)). After this transition, the matter asymmetries stay constant, while $B_{Y}$ decreases precisely exponentially and relatively slowly due to the expansion. The reason for the inclusion of Fig.\ 1e is to display clearly the changes of $B_{Y}$ for values below $10^{21}$ Gauss, and in particular show that at the end of the time interval, which is the onset of the electroweak phase transition, $B_{Y} \approx 10^{20}$. This final value is almost independent of its initial seed, as long as it is nonzero, and depends only on the initial matter asymmetries \cite{31,34}. Figure \ref{fig1} shows that by increasing the initial velocity, and hence the vorticity, the critical time decreases, or, equivalently, the critical temperature increases.

Next, we examine the behavior of the velocity field and the CVE more closely. First we should mention that since the hypermagnetic field is fully helical, i.e. $\vec{\nabla}\times\vec{B}_{Y}=\alpha\vec{ B}_{Y}$, it cannot affect the evolution of the velocity or the vorticity fields \cite{38}. This would, in the absence of viscosity, make the plasma force free. Indeed, these fields decrease exponentially due to the kinematic viscosity and rapidly tend to zero, as can be seen in Fig.\ \ref{fig1}. However, as stated earlier, their very brief presence can significantly affect the evolution of the hypermagnetic field and thus the matter asymmetries. 

\begin{figure}[!ht]
	\subfigure[]{\label{fig:figure:11}
		\includegraphics[width=.45\textwidth]{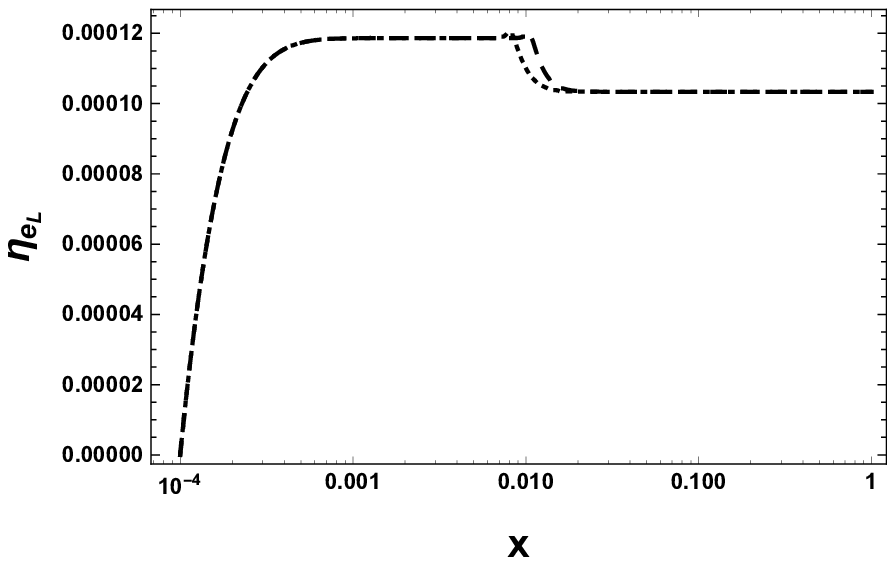}}
	\hspace{8mm}
	\subfigure[]{\label{fig:figure:21} 
		\includegraphics[width=.45\textwidth]{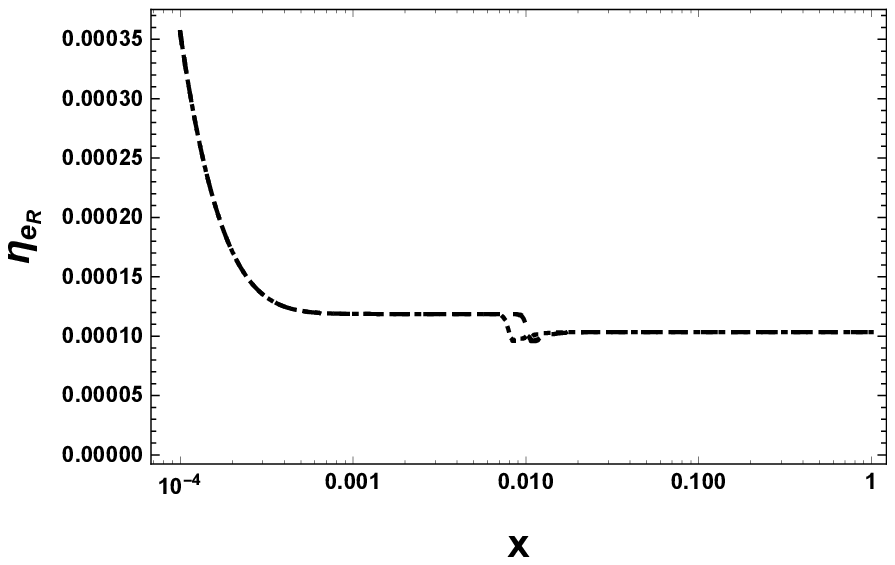}}
	\hspace{8mm}
	\subfigure[]{\label{fig:figure:31} 
		\includegraphics[width=.45\textwidth]{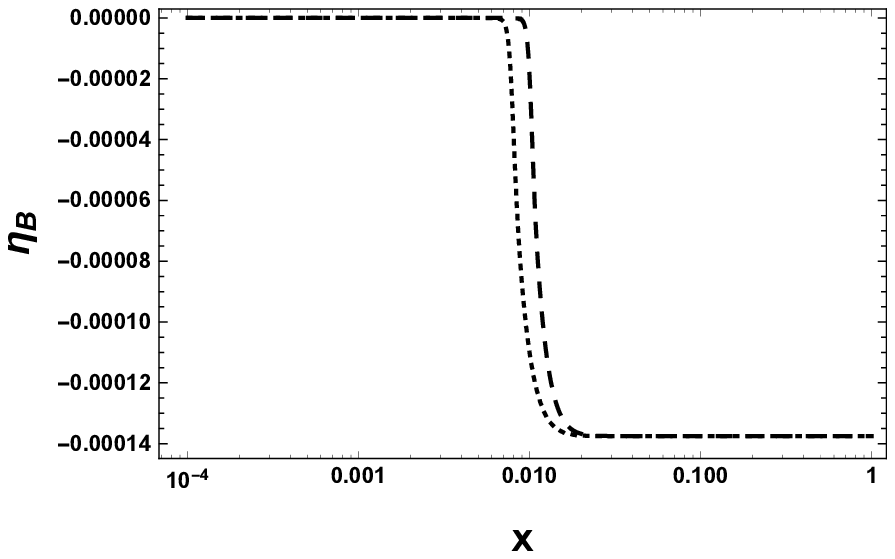}}
	\hspace{8mm}
	\subfigure[]{\label{fig:figure:41} 
		\includegraphics[width=.45\textwidth]{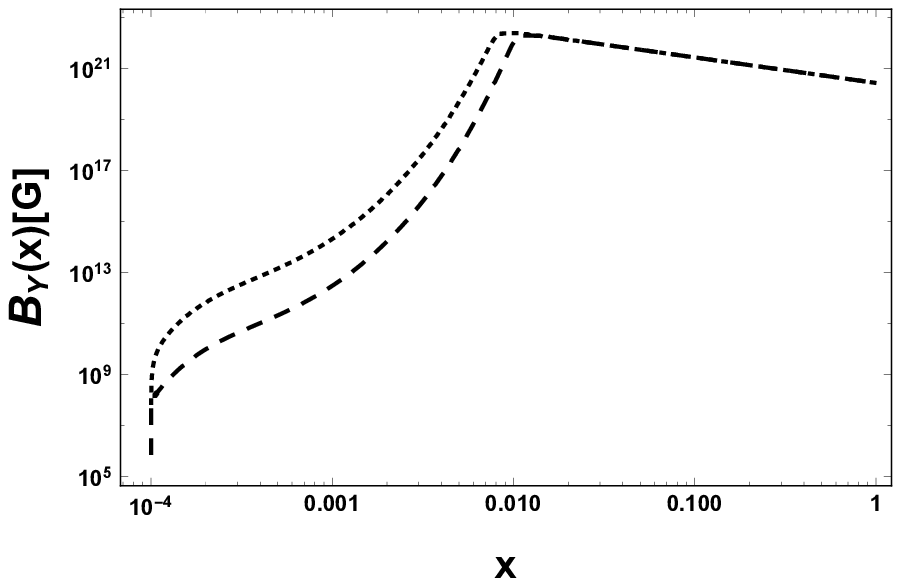}}
	\caption{\footnotesize Time plots of the lepton and the baryon asymmetries and the hypermagnetic field amplitude with the initial conditions $y_{R}^{(0)}=10^{3}$, $B_{Y}^{(0)}=0 $, and $y_{L}^{(0)}=y_{B}^{(0)}=0$, and $v_{0}=10^{-10}$. Dashed line is obtained for non-zero viscosity and dotted line for zero viscosity.\\  a: Left-handed lepton asymmetry, $\eta_{e_{L}}$. \\ b: Right-handed lepton asymmetry, $\eta_{e_{R}}$.\\ c: Baryon asymmetry, $\eta_{B}$. \\d: The amplitude of the hypermagnetic field, $B_{Y}$.}
	\label{fig4}
\end{figure}

The question that we address next is what would happen if the viscosity is zero. For this purpose
the set of coupled differential equations are solved with the initial conditions $y_{R}^{(0)}=10^{3}$, $B_{Y}^{(0)}=0$, $y_{L}^{(0)}=y_{B}^{(0)}=0$, and $v_{0}=10^{-10}$, in the presence and absence of viscosity. Figure \ref{fig4} shows that in a 
non-viscose plasma, although the velocity and the vorticity fields remain constant, the aforementioned effects due to the chiral vorticity on the matter asymmetries and the hypermagnetic field are not significantly altered. The most important effect of the absence of viscosity is that the seed produced for ${B}_{Y}$ by the vorticity is stronger. Hence, the CME can increase the amplitude of ${B}_{Y}$ to its saturation curve sooner, i.e., at higher value of critical temperature, as compared to the plasma with the non-zero viscosity. The drops in the values of matter asymmetries at the transition are unchanged.

This ineffective role of the vorticity after producing the initial seed for ${B}_{Y}$ is mainly due to the fact that after the electron chirality flip reactions come into equilibrium, the vorticity coefficient $c_v$ vanishes. This in turn is due to the fact that the contributions of the chemical potentials of the right-handed and the left-handed electrons to $c_v$ cancel each other. Therefore, after the electron chirality flip reactions come into equilibrium, the CVE is turned off, even if the vorticity is large. In fact, the chirality flip reactions in the temperature range under consideration are important and cannot be neglected. Finally,
it should be emphasized that the evolution of matter asymmetries and the hypermagnetic field amplitude are almost independent of the initial value of the vorticity and the viscosity, as long as the former is non-zero.

\section{CONCLUSION}\label{x5}
In this paper, we have studied the effects of the chiral vorticity on the evolution of the hypermagnetic field and the matter asymmetries in the early Universe and in the temperature range $100\mbox{ GeV} \le T\le 10\mbox{ TeV}$. Starting with an initial vorticity and large matter asymmetries at $10$ TeV, we have investigated the production and growth of the hypermagnetic field, and the evolution of the matter asymmetries and the vorticity till the onset of the electroweak phase transition, i.e. $100$ GeV. 

We have chosen the non-trivial Chern-Simons configuration with a monochromatic spectrum for the vector potentials of both the hypermagnetic and the velocity fields, with the same characteristic wave number $k=10^{-7}$ in the comoving frame. 
Since the hypermagnetic field is fully helical, i.e., $\vec{\nabla}\times\vec{B}=\alpha \vec{B}$, it has no effect on the evolution of the velocity or the vorticity fields. This is due to the fact that the term $\vec{J} \times\vec{ B}_{Y}$ vanishes in the Navier-Stokes equation. By considering an incompressible homogeneous plasma, the only remaining term in the Navier-Stokes equation is the kinematic viscosity, which leads to the exponential decrease of the vorticity.

Our most important result is that, an initial vorticity and matter asymmetries can produce a seed for the hypermagnetic field in the plasma via the CVE. This cannot occur if only the CME is taken into account. Subsequently, the CME leads to the growth of the hypermagnetic field amplitude until it reaches its maximum value. At this time a transition occurs where the matter asymmetries suddenly change, while preserving $B-L$, to attain their constant final values. We have shown that increasing the vorticity in the plasma leads to a stronger seed of the hypermagnetic field which then grows to yet a larger maximum value. Moreover, the critical time decreases, or equivalently the critical temperature increases. Later, the amplitude of the hypermagnetic field decreases gradually due to the expansion of the Universe, while its length scale increases as $\lambda=\frac{2\pi\sqrt{x}}{k T_{EW}}$, where $10^{-4}\le x\le1$. 

As mentioned before, if we choose the vector potentials of the hypermagnetic and velocity fields to be two different basis configurations, then their dot product in Eqs.\ (\ref{eq42},\ref{eq45}) would vanish. Therefore, The seed hypermagnetic field in Eq.\ (\ref{eq45}) will not be produced, and the subsequent evolution will be due only to the chirality flip processes, equalizing the chemical potentials of the right and left handed electrons. We have also investigated the case with the same initial matter asymmetry but with both vector potentials having the same negative helicity configuration. The result is that the generated $B_Y$ and $\eta_B$ are about 23 orders of magnitude smaller. For the cases in which the sign of initial matter asymmetries and helicities are simultaneously reversed, we obtain analogous evolutions but with the sign of asymmetries reversed. There are two generalization that can be considered for the vector potentials. First, one can consider fully helical configurations for the vector potentials, but with a superposition of wave numbers $k$. In this case we expect, due to the last term in Eq.\ (\ref{eq49}) which acts as a source term, the seed hypermagnetic field to still be generated. Second, one can include a non-helical component in the hypermagnetic field. In that case, $\vec{\nabla}\times\vec{B}_{Y}\neq\alpha\vec{ B}_{Y}$ and consequently the term $\vec{J} \times\vec{ B}_{Y}$ will not vanishes in the Navier-Stokes equation. This term acts as a source for vorticity and velocity, i.e. the plasma is no longer force free.

 \newpage
\section{APPENDIX A }\label{}
It is known that the Universe in large scale is homogeneous and isotropic, so its geometry can be described by a conformally flat metric of the Friedmann-Robertson-Walker (FRW) type with the form
\begin{equation}\label{eqa1}
 ds^{2}=dt^{2}-R^{2}(t)\delta_{ij}dx^{i}dx^{j},
\end{equation} 
where $t$ is the physical time, $x^{i}$s are the comoving coordinates, and $R(t)$ is the scale factor. Then, the effective Lagrangian density for the hypercharge gauge fields at finite fermion density and in the curved space-time can be written as \cite{{volovik},{volovik2}}
 \begin{equation}\label{eqa2}
 \begin{split}
 \mathsterling&=\sqrt{-g}\hat{\mathsterling}\\&=\sqrt{-g}\left[-\frac{1}{4}F_{\mu\nu}F^{\mu\nu}-
 {J}^{\mu}_{\mathrm{Ohm}}A_{\mu} + \frac{c_B}{4}\tilde{\epsilon}_{ijk}F^{ij}A^{k}R^{3}+ \frac{c_{\mathrm{v}}}{2}\tilde{\epsilon}_{ijk}\omega^{ij}A^{k}R^{3}\right],
 \end{split}
 \end{equation}
where $F^{\mu\nu}=\nabla^{\mu}A^{\nu}-\nabla^{\nu}A^{\mu}$ is the field strength tensor, $A_{\mu}$ is the hypercharge vector potential, $g$ is the  determinant of the FRW metric defined in Eq.\ (\ref{eqa1}), and $\nabla_{\mu}$ is the covariant derivative with respect to this metric. Moreover, ${J}^{\mu}_{\mathrm{Ohm}}=(J^{0},\vec{J}/R)$ is the Ohmic four-vector current, $\tilde{\epsilon}_{ijk}=-\tilde{\epsilon}^{ijk}$ is the Levi-Civita symbol, $\omega^{ij}=\nabla^{i}u^{j}-\nabla^{j}u^{i}$ is the antisymmetric vorticity tensor, and $\vec{u}=\vec{v}/R$ is the bulk velocity of the plasma in the curved space-time. The vorticity and the helicity coefficients $c_{\mathrm{v}}$ and $c_{B}$ appearing in Eq.\ (\ref{eqa2}) are given in Eqs.\ (\ref{eq22},\ref{eq25}), respectively. Using the effective Lagrangian density, as given by Eq.\ (\ref{eqa2}), in the following equation 
 \begin{equation}\label{eqa3}
 \frac{\partial \hat{\mathsterling}}{\partial A_{\nu}}-\nabla_{\mu}\left[\frac{\partial \hat{\mathsterling}}{\nabla_{\mu} A_{\nu}}\right]=0,
 \end{equation}
the Euler-Lagrange equations for the hypercharge gauge fields in the curved space-time can be obtained as 
\begin{equation}\label{eqa4}
 \nabla_{\mu}F^{\mu\nu}={J}^{\nu}_{\mathrm{Ohm}}-\frac{c_{B}}{2}\tilde{\epsilon}_{ijk}F^{ij}g^{kk}\delta^{\nu}_{k} R^{3}(t)-\frac{1}{R^{2}(t)} c_{\mathrm{v}}\tilde{\epsilon}_{ijk}\left(\nabla_{i}v^{j}\right)\delta^{\nu}_{k}.
 \end{equation}
Note also that the only non-vanishing Christoffel symbols of the metric (\ref{eqa1}) are $\Gamma^{0}_{ij}=R\dot{R}\delta_{ij}$ and $\Gamma^{i}_{0j}=\Gamma^{i}_{j0}=\dot{R}/R\delta^{i}_{j}$. It can be seen that three different types of electric current appear on the rhs of Eq.\ (\ref{eqa4}). These are the Ohmic current ${J}^{\nu}_{\mathrm{Ohm}}$, the zeroth component of which is zero due to the hypercharge neutrality in the plasma, the chiral magnetic current $J^{\nu}_{\mathrm{cm}}=\left(0,c_{B}\tilde{\epsilon}_{ijk}F^{ij}R/2\right)$, and the chiral vortical current $J^{\nu}_{\mathrm{cv}}=\left(0,-c_{\mathrm{v}}\tilde{\epsilon}_{ijk}\left(\nabla_{i}v^{j}\right)/R^{2}\right)$. By using $F^{ij}=-\tilde{\epsilon}^{ijk}\left(B^{k}/R^{2}\right)$ and $\tilde{\epsilon}_{ijk}(\nabla_{i}v^{j})=w_{k}=-w^{k}$, these chiral currents simplify to $J^{\nu}_{\mathrm{cm}}=(0,c_{B}\vec{B}_{Y}/R)$ and $J^{\nu}_{\mathrm{cv}}=\left(0,c_{\mathrm{v}}\vec{w}/R\right)$, respectively. 

Considering $\nu=0$ in Eq.\ (\ref{eqa4}), the Gauss's Law is obtained as
 \begin{equation}\label{eq4a}
 \frac{1}{a}\vec{\nabla}.\vec{E}_{Y}=\rho_{\mathrm{total}}=0,
 \end{equation}
the rhs of which vanishes due to the hypercharge neutrality of the plasma. Then, considering $\nu=i$ in Eq.\ (\ref{eqa4}), 
the time evolution of the hyperelectric field in the presence of the CME and the CVE, and in the expanding Universe (Ampere's Law) will be obtained as
 \begin{equation}\label{eq4a1}
 \partial_{t}\vec{E}_{Y}+2H\vec{E}_{Y}=\frac{1}{R}\left(\vec{\nabla}\times\vec{B}_{Y}\right)-\vec{J}_{\mathrm{Ohm}}-c_{B}\vec{B_{Y}}-c_{\mathrm{v}}\vec{\omega}.
 \end{equation}
In the above equation, the term $2H\vec{E}_{Y}$ is due to the scaling of the hyperelectric field in the expanding Universe.
In order to obtain the two other Maxwell's equations, the following Bianchi identity is used
 
 \begin{equation}\label{equ4}
 \nabla_{\mu}F_{\nu\rho}+\nabla_{\rho}F_{\mu\nu}+\nabla_{\nu}F_{\rho\mu}= \partial_{\mu}F_{\nu\rho}+\partial_{\rho}F_{\mu\nu}+\partial_{\nu}F_{\rho\mu}=0,
 \end{equation}
 which results in
 \begin{equation}\label{equu4}
 \vec{\nabla}.\vec{B}_{Y}=0,
 \end{equation}
 and
 \begin{equation}\label{equuu4}
 \partial_{t}\vec{B}_{Y}+2H\vec{B}_{Y}=-\frac{1}{R}\left(\vec{\nabla}\times\vec{E}_{Y}\right).
 \end{equation}
It can be seen that, similar to the hyperelectric field, the hypermagnetic field is also scaled as $R^{-2}$.
 
\section{APPENDIX B}
The plasma of the early Universe contains different types of constituents which are sufficiently strongly coupled to be considered as a fluid \cite{6}.
Moreover, It can be considered as an ideal fluid with the equation of state $p=\rho/3$ in the radiation dominated era, where $p$ and $\rho$ are the pressure and the energy density of the plasma, respectively.
The energy momentum tensor of this ideal fluid in the presence of the hypercharge electromagnetic fields can be written as
 \begin{equation}\label{eq234}
 T^{\mu\nu}=T^{\mu\nu}_{f}+T^{\mu\nu}_{\mathrm{em}},
 \end{equation}
 where
 \begin{equation}\label{eq411}
T^{\mu\nu}_{f}= (\rho+p)U^{\mu}U^{\nu}- p g^{\mu\nu},
 \end{equation}
 and
 \begin{equation}\label{eq124}
 T^{\mu\nu}_{\mathrm{em}}=\frac{1}{4}g^{\mu\nu} F^{\alpha\beta} F_{\alpha\beta}-F^{\nu\sigma}{F^{\mu}}_{\sigma}.
 \end{equation}
In the above equations, $F_{\alpha\beta}=\nabla_{\alpha}A_{\beta}-\nabla_{\beta}A_{\alpha}$, $U^{\mu}=\gamma\left(1,\vec{v}/R\right)$ is the four-velocity of the plasma normalized such that $U^{\mu}U_{\mu}=1$, and $\gamma$ is the Lorentz factor. 
Due to the ideal fluid assumption, the non-ideal effects are ignored in Eq.\ (\ref{eq411}) \cite{b1234}.  
Since the Einstein tensor obtained from the metric (\ref{eqa1}) is diagonal, not only the hypercharge electromagnetic field density must be small compared to the energy density of the Universe \cite{b1234}, but also the bulk velocity should respect the condition $\left |\vec{v}\right | \ll1$, or equivalently $\gamma\simeq1$ and $U^{\mu}\simeq(1,\vec{v}/R)$. Using the conservation equation of the energy momentum tensor $\nabla_{\mu}T^{\mu\nu}=0$, the conservation equation of the energy density and the continuity equation can be obtained. Considering $\nu=0$, we obtain 

 \begin{equation}\label{eq432}
 \partial_{t}\rho+ \vec{\nabla}.\left[(\rho+p)\frac{\vec{v}}{R}\right]+3H(\rho+p)\left(1+v^{2}\right)=\vec{E}_{Y}.\vec{J},
\end{equation}
where $\vec{J}=\vec{J}_{\mathrm{Ohm}}+c                                                                                                                                                                                                                                            _{B}\vec{B}_{Y}+c_{\mathrm{v}}\vec{\omega}$. 
The second order term in the velocity field and the term $\vec{E}_{Y}.\vec{J}$ appearing in the above equation are usually neglected. 
Considering $\nu=j$, the continuity equation can be obtained as

\begin{equation}\label{eq4aq}
\begin{split} 
&\left[\partial_{t}\rho+ \frac{1}{R}\vec{\nabla}.\left[(\rho+p)\vec{v}\right]+3H(\rho+p)\right]\vec{v}+\left[\partial_{t}p+H(\rho+p)\right]\vec{v}\\&+(\rho+p)\partial_{t}\vec{v}+(\rho+p)\frac{\vec{v}.\vec{\nabla}}{R}\vec{v}+\frac{\vec{\nabla} p}{R}=
\rho_{\mathrm{total}}\vec{E}_{Y}-\left(\vec{B}_{Y}\times\vec{J}_{\mathrm{Ohm}}\right)- c_{\mathrm{v}}\vec{B}_{Y}\times\vec{\omega}.
\end{split} 
\end{equation}
On the rhs of Eq.\ (\ref{eq4aq}), the second and the third terms are obtained from $\vec{B}_{Y}\times\vec{J}$. Furthermore, the term $\rho_{\mathrm{total}}\vec{E}_{Y}$ vanishes since $J^{0}_{\mathrm{cm}}=J^{0}_{\mathrm{cv}}=0$ and the plasma is electrically neutral. 

Let us now obtain the equations for the anomalous divergence of the matter currents in the symmetric phase and in the curved space-time.
Due to the chiral coupling of the hypercharge fields to the fermions, the fermion numbers are violated as
\begin{equation}\label{eq4aq1}
\nabla_{\mu}J^{\mu}_{i}=C_{i} \vec{E}_{Y}.\vec{B}_{Y},
\end{equation} 
where $J^{\mu}_{i}$ is the fermionic current and $C_{i}$ is its corresponding Anomaly coefficient. The above equation can also be written in the form

\begin{equation}\label{eq4aq2}
\partial_{t}J^{0}_{i}+ \frac{1}{R}\vec{\nabla}.\vec{J}_{i}+ 3HJ^{0}_{i} =C_{i} \vec{E}_{Y}.\vec{B}_{Y}.
\end{equation} 
Then, by integrating over all space and dividing by volume, the second term vanishes and we obtain
  \begin{equation}\label{eq4aq3}
 \partial_{t}(n_{i}-\bar{n}_{i})+  3H\left(n_{i}-\bar{n}_{i}\right) =C_{i} \langle\vec{E}_{Y}.\vec{B}_{Y}\rangle,
 \end{equation}
where $n_{i}$ and $\bar{n}_{i}$ are the number densities of the ith species of the fermion and the anti-fermion, respectively.
Using the relation $\dot{s}/s=-3H$, we obtain
 \begin{equation}\label{eq4aq3}
s \partial_{t}\left(\frac{n_{i}-\bar{n}_{i}}{s}\right)=C_{i} \langle\vec{E}_{Y}.\vec{B}_{Y}\rangle,
 \end{equation}
where $s$ is the entropy density.

\end{document}